\RequirePackage{fix-cm}
\documentclass[smallextended]{svjour3}       % onecolumn (second format)
\smartqed  % flush right qed marks, e.g. at end of proof
\usepackage{graphicx}
\usepackage{amssymb}

\def\NH#1#2{^{#2}{\rm #1}}
\def\SH#1#2{^{#2}_{\Lambda}{\rm #1}}
\def\Lnn{\Lambda {\rm nn}}
\def\DH#1#2{^{#2}_{\Lambda\Lambda}{\rm #1}}

\begin{document}

\title{Novel method for producing very-neutron-rich hypernuclei via charge-exchange reactions with heavy ion projectiles}


\author{Takehiko R. Saito         \and
Hiroyuki Ekawa \and
Manami Nakagawa
}


\institute{Takehiko. R. Saito \at
              High Energy Nuclear Physics Laboratory, Cluster for Pioneering Research, RIKEN, 2-1 Hirosawa, Wako, Saitama 351-0198, Japan \\
              GSI Helmholtz Centre for Heavy Ion Research, Planckstrasse 1, D-64291 Darmstadt, Germany\\
              School of Nuclear Science and Technology, Lanzhou University, 222 South Tianshui Road, Lanzhou, Gansu Province, 730000, China\\
              \email{takehiko.saito@riken.jp}           %  \\
           \and
           Hiroyuki Ekawa  \at
            High Energy Nuclear Physics Laboratory, Cluster for Pioneering Research, RIKEN, 2-1 Hirosawa, Wako, Saitama 351-0198, Japan
            \and 
            Manami Nakagawa \at
            High Energy Nuclear Physics Laboratory, Cluster for Pioneering Research, RIKEN, 2-1 Hirosawa, Wako, Saitama 351-0198, Japan
}

\date{Received: date / Accepted: date}

\maketitle

\begin{abstract}
We propose a novel method for producing very-neutron-rich hypernuclei and corresponding resonance states by employing charge-exchange reactions via pp($^{12}$C, $^{12}$N $K^+$)n$\Lambda$ with single-charge-exchange and ppp($^{9}$Be, $^{9}$C $K^+$)nn$\Lambda$ with double-charge-exchange, both of which produce $\Lambda K^+$ in a target nucleus. The feasibility of producing very-neutron-rich hypernuclei using the proposed method was analysed by applying an ultra-relativistic quantum molecular dynamics model to a $^6$Li+$^{12}$C reaction at 2 $A$ GeV. The yields of very-neutron-rich hypernuclei, signal-to-background ratios, and background contributions were  investigated. The proposed method is a powerful tool for studying very-neutron-rich hypernuclei and resonance states with a hyperon for experiments employing the Super-FRS facility at FAIR and HFRS facility at HIAF. 

\end{abstract}

%%%%%%%%%%%%%%%%%%%%%%%%%%%%%%%%%%%%%%%%%%%%%%%%%%%%%%%%%%%%%%
\section{Introduction}

Hypernuclei, which are sub-atomic bound states with at least one hyperon, have been studied for approximately seven decades, mainly by employing induced reactions of cosmic rays, secondary meson-beams, and primary electron beams \cite{DAVIS20053}. These investigations of hypernuclei have extended our understanding of nuclear forces under the flavoured-SU(3) symmetry. In the past decade, experimental studies of hypernuclei have entered a new era based on hypernuclear spectroscopy with heavy ion projectiles bombarded onto a fixed nuclear target. The HypHI Phase 0 experiment at GSI successfully produced and observed light hypernuclei with
$ \SH{H}{3}$, $\SH{H}{4}$, and $\Lambda$-hyperons by employing $^6$Li+$^{12}$C reactions at 2 $A$ GeV \cite{RAPPOLD2013170}. In this experiment, hypernuclei were produced as projectile fragments and observed during flight. Based on the nature of fragmentation reactions of heavy ion beams, the isospin values of the produced hypernuclei were widely distributed. Therefore, neutron-rich and proton-rich hypernuclei could be studied. 

One of the problems revealed by the results of the HypHI Phase 0 experiment is the possible existence of an unprecedented bound state of a $\Lambda$-hyperon with two neutrons, denoted as $\Lnn$ ($\SH{n}{3}$) \cite{PhysRevC.88.041001}. Neutral nuclear states with neutrons and $\Lambda$-hyperons are of particular interest because the natures of these states should have an impact on our understanding of the deep cores of neutron stars. However, theoretical calculations have shown negative results for the existence of $\Lnn$ bound states \cite{PhysRevC.89.061302,GAL201493,PhysRevC.89.057001,SCHAFER2020135614}. Although there is disagreement between the results of the HypHI Phase 0 experiment and theoretical calculations, whether or not the $\Lnn$ state can exist has recently become a hot topic in experimental and theoretical nuclear physics. The possible resonance states of $\Lnn$ have also been studied theoretically \cite{SCHAFER2020135614,PhysRevC.92.054608}. However, it should be noted that such $\Lnn$ resonance states could not be observed in the HypHI Phase 0 experiment because of the displaced vertex trigger used in its data acquisition system \cite{RAPPOLD2013170}.
Additionally, interest in neutral hypernuclei such as $\Lnn$ has grown. A recent reanalysis of the BNL-AGS E906 experiment revealed the possibility of the existence of a bound double-strangeness neutral hypernucleus, denoted as $\DH{n}{4}$ \cite{BLESER2019502}. Additional information regarding neutral hypernuclei and resonance states with a hyperon (hereafter called `hyper-resonance states'), including $\Lnn$ and states with more neutrons, is required to understand the equation of state in the interiors of neutron stars. However, the production and observation of such neutral states using existing experimental methods are extremely difficult. 

As described above, hypernuclear production with heavy ion beams, which was employed in the HypHI Phase 0 experiment, can be a useful tool for studying neutron-rich hypernuclei. However, it is not always practical to use this method for studying very-neutron-rich hypernuclei. The probability of producing very-neutron-rich hypernuclei via neutral or very-neutron-rich projectile fragments capturing a $\Lambda$-hyperon is very small. Additionally, the decay of neutral or very-neutron-rich hypernuclei emits several neutrons, which requires simultaneous multi-neutron measurements to reconstruct the produced hypernuclei using the invariant mass method. Simultaneous multi-neutron measurements could be made possible with modern neutron detectors such as the NeuLAND detector at FAIR \cite{NeuLAND}, but the required accuracy and confidence level for studying those states are unachievable. 

In this article, we propose a novel method for producing very-neutron-rich hypernuclei through the single- and double-charge-exchange reactions of heavy ion beams combined with a conversion process of Np$\rightarrow $N$\Lambda K^+$. In this method, one or two protons in a target nucleus are converted into neutrons by charge-exchange reactions and an additional proton in the target nucleus is converted into a $\Lambda$-hyperon. Therefore, very-neutron-rich hypernuclei can be produced. Additionally, one or two neutrons in a projectile can be converted into protons by charge-exchange reactions, which can be analysed by a forward spectrometer. The produced neutron-rich hypernuclei can be observed using the missing mass method based on coincident measurements of outgoing $K^+$ and charge-exchanged projectile residue. 

The proposed reaction and method for producing very-neutron-rich hypernuclei are presented in Section \ref{sec:overview}. Section \ref{sec:cal} discusses calculations using a microscopic transport model to study the production of very-neutron-rich hypernuclei using charge-exchange reactions. Section \ref{sec:discussion} discusses the results of our calculations and the feasibility of the proposed method.

\begin{table}[pt]
\caption{Summary of hypernuclei/resonances and proposed charge-exchange reactions for $Z=0\sim 8$.}
\label{tbl:list}
\begin{center}
\begin{tabular}{ c | c | c | c | c}
\hline \hline
 & Single-charge  & Double-charge   & Produced  & Former\\
 Target & exchange  & exchange  & hypernuclei & observation\\
 & ($^{12}$C, $^{12}$N $K^+$) & ($^{9}$Be, $^{9}$C $K^+$) & or resonance& \\
\hline
$^3$He &	\checkmark   &		&  $\SH{n}{3}$ ($\Lnn$) & \cite{PhysRevC.88.041001}\\
$^4$He &	\checkmark   &		&  $\SH{n}{4}$  & \\
$^6$Li &  &	\checkmark	&  $\SH{n}{6}$  & \\
$^7$Li &  &	\checkmark	&  $\SH{n}{7}$  & \\
\hline
$^6$Li &	\checkmark   &		&  $\SH{H}{6}$  & \cite{PhysRevLett.108.042501}\\
$^7$Li &	\checkmark   &		&  $\SH{H}{7}$  & \\
$^9$Be &	   &	\checkmark	&  $\SH{H}{9}$  & \\
\hline
$^9$Be &	\checkmark   &		&  $\SH{He}{9}$  & \\
$^{10}$B &	   &	\checkmark	&  $\SH{He}{10}$  & \\
\hline
$^{10}$B &	\checkmark   &		&  $\SH{Li}{10}$  & \cite{PhysRevLett.94.052502}\\
$^{11}$B &	\checkmark   &		&  $\SH{Li}{11}$  & \\
$^{12}$C &	   &	\checkmark	&  $\SH{Li}{12}$  & \\
\hline
$^{12}$C &	\checkmark   &		&  $\SH{Be}{12}$  & \\
$^{14}$N &	   &	\checkmark	&  $\SH{Be}{14}$  & \\
\hline
$^{14}$N &	\checkmark   &		&  $\SH{B}{14}$  & \\
$^{16}$O &	   &	\checkmark	&  $\SH{B}{16}$  & \\
\hline
$^{16}$O &	\checkmark   &		&  $\SH{C}{16}$  & \\
$^{19}$F &	   &	\checkmark	&  $\SH{C}{19}$  & \\
\hline
$^{19}$F &	\checkmark   &		&  $\SH{N}{19}$  & \\
$^{20}$Ne &	   &	\checkmark	&  $\SH{N}{20}$  & \\
\hline
$^{20}$Ne &	\checkmark   &		&  $\SH{O}{20}$  & \\
$^{23}$Na &	   &	\checkmark	&  $\SH{O}{23}$  & \\
\hline \hline
\end{tabular}
\end{center}
\end{table}

%%%%%%%%%%%%%%%%%%%%%%%%%%%%%%%%%%%%%%%%%%%%%%%%%%%%%%%%%%%%%%
\section{Proposed hypernuclear production method using single- and double-charge-exchange reactions}
\label{sec:overview}

Singe- and double-charge-exchange reactions of heavy ion beams have already been used as powerful tools for studying exotic nuclei. One important result of such reactions is the production and observation of a tetra-neutron resonance state using a double-charge-exchange reaction of $^4$He($^8$He, $^8$Be)4n at 186 $A$ MeV \cite{PhysRevLett.116.052501}. To produce hypernuclei using induced reactions of heavy ion beams, such as charge-exchange reactions, the energy of the heavy ion beams should be greater than the energy threshold required to produce a $\Lambda$-hyperon, which is approximately 1.7 GeV. The beam energy employed in the HypHI Phase 0 experiment was 2 $A$ GeV \cite{RAPPOLD2013170}. However, data on charge-exchange reactions in this energy regime are scarce. Additionally, hypernuclei have never been studied based on the charge-exchange reactions of heavy ion beams. Although the feasibility of the production of hypernuclei has not yet been experimentally demonstrated, we consider that charge-exchange reactions could be a powerful tool for producing and studying very-neutron-rich hypernuclei. Therefore, we investigated the feasibility of the proposed production method using events produced by a microscopic transport model.   

We propose a novel method with single- and double-charge-exchange reactions at 2 $A$ GeV or higher to convert target nuclei into very-neutron-rich hypernuclei or hyper-resonance states. In the single-charge-exchange reaction, one of the protons in the target nucleus is converted into a neutron, while one of the neutrons in the projectile nucleus is converted into a proton. For the double-charge-exchange reactions, two protons in the target nucleus become two neutrons and two protons in the projectile are converted into two neutrons. Additionally, in the proposed method, a $\Lambda$-hyperon and $K^+$ meson are produced from one of the protons in the target nucleus via collision with the projectile. 
As a result of these reactions, two or three protons in the target nucleus are converted into n$\Lambda$ or nn$\Lambda$ by single- or double-charge-exchange reactions with $\Lambda K^+$ production, respectively, and very-neutron-rich hypernuclei and hyper-resonances can be produced.  
We wish to observe produced very-neutron-rich hypernuclei or hyper-resonances using a simple missing mass method by measuring the four-momentum of incident heavy ion beams, $K^+$, and outgoing projectile residue. Therefore, we require that the outgoing projectile residue have the same mass number as the incident projectile, but with a different charge number caused by the charge-exchange reaction, to ensure that there are no additional nucleons emitted from the projectile residue. This can be achieved by performing peripheral gentle collisions, which provide a reasonable probability of the production of target residues with the same mass number as the target nucleus without emitting neutrons.     

One important consideration for the choice of outgoing projectile residues is that they should not have any excited states to avoid ambiguity in missing mass measurements. Additionally, we prefer to use primary beams from accelerators because their energy and momenta are very well defined. In projectile residues, there are two good nuclei $^9$C and $^{12}$N, which are bound only by their ground states. These residues can be produced by stable primary beams $^9$Be and $^{12}$C via double- and single-charge-exchange reactions, (i.e., ($^{9}$Be, $^{9}$C), and ($^{12}$C, $^{12}$N)). By combining these charge-exchange reactions with $\Lambda K^+$ production in the target, several protons in the target nucleus are converted into neutrons and a $\Lambda$-hyperon, such as pp($^{12}$C, $^{12}$N $K^+$)n$\Lambda$ and ppp($^{9}$Be, $^{9}$C $K^+$)nn$\Lambda$. 

Table \ref{tbl:list} summarises the possible productions of neutral and very-neutron-rich hypernuclei, as well as the hyper-resonance states from $Z=0$ up to $Z=8$, when using the proposed single- and double-charge-exchange reactions with $\Lambda K^+$ production. As shown in the table, the existence of the $\Lnn$ bound state can be studied based on the $^3$He($^{12}$C, $^{12}$N $K^+$) reaction. Resonance states involving $\Lnn$ can also be studied in the same reaction. A very-neutron-rich hydrogen hypernucleus denoted as $\SH{H}{6}$ was first discovered by the FINUDA collaboration \cite{PhysRevLett.108.042501}. However, it was not observed in the E10 experiment at J-PARC \cite{201439}. The existence of the $\SH{H}{6}$ hypernucleus can be studied and confirmed in single-charge-exchange reactions with a $^6$Li target, namely $^6$Li($^{12}$C, $^{12}$N $K^+$)$\SH{H}{6}$. A neutron-rich lithium hypernucleus denoted as $\SH{Li}{10}$ was observed in the KEK-PS-E521 experiment, but with small statistics \cite{PhysRevLett.94.052502}. This hypernucleus can also be studied with larger statistics by employing the single-charge-exchange reaction $^{10}$B($^{12}$C, $^{12}$N $K^+$)$\SH{Li}{10}$. The other hypernuclei and hyper-resonance states listed in Table \ref{tbl:list} have never been observed or studied, but the reaction proposed in this paper will provide a wide variety of opportunities to study these states. The proposed novel method can be applied using the Super-FRS \cite{Super-FRS} facility at FAIR \cite{FAIR} or HFRS facility at HIAF \cite{HIAF}. A similar detector system from the WASA-FRS hypernuclear experiment \cite{WASA.FRS} can also be implemented in the mid-focal plane of the Super-FRS and HFRS. 

As discussed above, the novel method proposed in this paper could open new avenues for studying neutron-rich hypernuclei and hyper-resonances. However, it is expected that the cross section of the charge-exchange reactions with $\Lambda K^+$ production in the target nucleus will be small and that the background induced by the breakup of target residues and pion  productionwill be significant. Based on a lack of experimental data on charge-exchange reactions at 2 $A$ GeV or higher and a lack of data on these reactions in coincidence with $\Lambda K^+$  production, we performed feasibility studies on the production of neutron-rich hypernuclei and hyper-resonances by employing an ultra-relativistic quantum molecular dynamics model called UrQMD \cite{BASS1998255,Bleicher_1999}, which will be discussed in the next section.

%%%%%%%%%%%%%%%%%%%%%%%%%%%%%%%%%%%%%%%%%%%%%%%%%%%%%%%%%%%%%%
\section{Production of very-neutron-rich hypernuclei with UrQMD in a $^6$Li+$^{12}$C reaction at 2 $A$ GeV}
\label{sec:cal}

For our feasibility studies on the production and identification of very-neutron-rich hypernuclei and hyper-resonance states by means of charge-exchange reactions combined with $\Lambda K^+$ production in the target nucleus, we employed an ultra-relativistic quantum molecular dynamics model called UrQMD \cite{BASS1998255,Bleicher_1999}. This model was developed to simulate relativistic heavy ion collisions in the energy range of approximately 1 GeV (energy of SIS at GSI) to the centre-of-mass energy of a few hundred GeV per nucleon pair at RHIC. It was also used as an event generator for Monte Carlo simulations in the HypHI Phase 0 experiment \cite{RAPPOLD2013170}. The UrQMD model includes charge exchanges between projectiles and target nuclei with elastic and inelastic scattering processes. 

In this work, we employed events generated by UrQMD for the reaction of $^6$Li+$^{12}$C at 2 $A$ GeV and performed clustering to form target and projectile fragments, as well as hypernuclei, in the target- and projectile-rapidity regions based on a kinematic cut for produced baryons in the UrQMD events. Although the proposed reactions involve $^9$Be and $^{12}$C projectiles impinged on different target nuclei, this reaction was selected because clustering for producing fragments and hypernuclei has already been well studied and optimised in this reaction for the HypHI Phase 0 experiment, which will be discussed later. 
This study was performed with approximately 15 billion pre-produced UrQMD events and used for Monte Carlo simulations of the WASA-FRS hypernuclear experiment in FAIR Phase 0 \cite{WASA.FRS}. As discussed later in this article, the corresponding statistics are not sufficiently large to reveal the details of perfect cases for producing hypernuclei with charge-exchange reactions and $\Lambda K^+$ production. However, feasibility can be discussed by using various channels within the 15 billion UrQMD events. In the future, the details of the proposed method will be analysed using much larger statistics, as well as experimental studies using beams from accelerators.   

\begin{figure*}[htb]
\centering
  \resizebox{120mm}{!}{\includegraphics{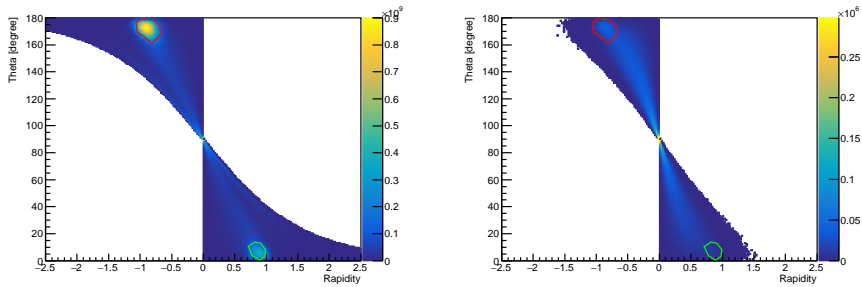}} 
\caption{Distributions of angles relative to the beam direction in degrees in the laboratory frame as a function of rapidity for nucleons (left panel) and $\Lambda$-hyperons (right panel). Polygons with red and blue colours represent the current conditions for forming target and projectile fragments, respectively. }
\label{fig:theta_rapidity}
\end{figure*}

The formation of nuclear fragments and hypernuclei in target- and projectile-rapidity regions was performed by considering the kinematics of the nucleons and hyperons. The left panel in Fig. \ref{fig:theta_rapidity} presents distributions of angles relative to the beam direction as a function of rapidity for all produced particles in a nucleon-nucleon centre-of-mass system. The dense regions around rapidity $=-0.9$ and $0.9$ correspond to the formation of target and projectile fragments with nucleons, respectively. For each event, when a pair of (rapidity, angle) of nucleons is inside the polygons indicated by red and green colours in the figure, they participate in the formation of target- and projectile-fragments as bound states, respectively. For the fragments, the kinematics of all nucleons inside the polygons is considered. The right panel in Fig. \ref{fig:theta_rapidity} presents a similar distribution for produced $\Lambda$-hyperons and one can see that the rapidity is widely distributed around rapidity $=0$. For each event, when the (rapidity, angle) pair of a $\Lambda$-hyperon is inside the red and green polygons, it participates with other nucleons inside the polygons to form hypernuclei at target- and projectile-rapidity regions, respectively (assuming that all hypernuclei are bound).     

\begin{figure*}[htb]
\centering
  \resizebox{120mm}{!}{\includegraphics{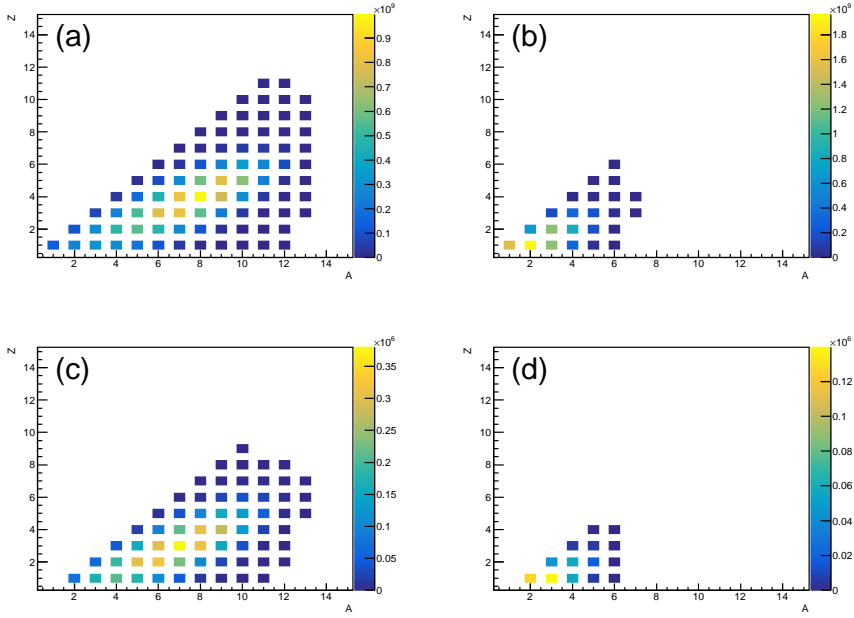}} 
\caption{(a) Correlations between the mass number ($A$) and charge ($Z$) for produced nuclear fragments in the target-rapidity region, (b) nucleon fragments in the projectile-rapidity region, (c) $\Lambda$-hypernuclei in the target-rapidity region, and (d) $\Lambda$-hypernuclei in the projectile-rapidity region.}
\label{fig:AZ_all}
\end{figure*}

Fig. \ref{fig:AZ_all} presents the correlations between the mass number ($A$) and charge ($Z$) for produced nuclear fragments (with nucleons) in the target-rapidity region (panel (a)), nuclear fragments in the projectile-rapidity region (panel (b)), $\Lambda$-hypernuclei in the target-rapidity region (panel (c)), and $\Lambda$-hypernuclei in the projectile-rapidity region (panel (d)). One can see that a variety of nuclear fragments and $\Lambda$-hypernuclei are produced in both the target- and projectile-rapidity regions. To evaluate the formation of nuclear fragments and hypernuclei, the production of the $\SH{H}{3}$ hypernucleus in this model was compared to the data obtained from the HypHI Phase 0 experiment. In our calculations using UrQMD, a total of 139525 $\SH{H}{3}$ hypernuclei were produced in the projectile-rapidity region, as shown at $A=3$ and $Z=1$ in panel (d), by the total of $1.498\times 10^{10}$ produced UrQMD events. The production cross section of $\SH{H}{3}$ in the projectile-rapidity region can be roughly estimated using the following relationship: 
\begin{equation}
\frac{\sigma }{A_{\rm reaction}} = \frac{N}{N_{\rm total }}, 
\label{eqn:eqn1}
\end{equation}
where $\sigma$ denotes the cross section of the products of interest and $A_{\rm reaction}$ is the area of the reaction occurring in the UrQMD calculations. In this work, an impact parameter was randomly selected between 0 and 3.5 fm for every event in the UrQMD calculations. Therefore, $A_{\rm reaction} = \pi \times 3.5 ^2$ fm$^2$. On the right side of the equation, $N_{\rm total }$ is the total event number, which is $1.498\times 10^{10}$, and $N$ denotes the number of products of interest. For the production of $\SH{H}{3}$ hypernuclei in the projectile-rapidity region, $N=139525$.  With these values, the cross section required to produce $\SH{H}{3}$ in the projectile-rapidity region was determined to be 3.6 $\mu $b. This cross section value was compared to the measured cross section of $\SH{H}{3}$ in the projectile-rapidity region with $^6$Li+$^{12}$C at 2 $A$ GeV in the HypHI Phase 0 experiment, which is $3.9\pm 1.4$ $\mu $b \cite{RAPPOLD2015129}. These values are reasonably close. The production of $\Lambda$-hyperons integrated over all rapidity regions was also evaluated by using Eq. (\ref{eqn:eqn1}) with $N=7.70 \times 10^7$. The derived cross section for $\Lambda$-hyperon production is 1.98 mb. This value was also compared to the measured cross section value of $1.7\pm 0.8$ mb in the HypHI Phase 0 experiment \cite{RAPPOLD2015129}. Again, there is reasonable agreement between the derived and measured values. These agreements indicate that the models used in this work for the formation of hypernuclei and $\Lambda$-hyperons are reasonable. Therefore, we used this model to study the production of very-neutron-rich hypernuclei. 

In the next section, the feasibility of hypernuclear production using single- and double-charge-exchange reactions combined with $\Lambda K^+$ production in the target-rapidity region will be discussed based on the nuclear fragments and hypernuclear produced in the UrQMD calculations.

%%%%%%%%%%%%%%%%%%%%%%%%%%%%%%%%%%%%%%%%%%%%%%%%%%%%%%%%%%%%%%
\section{Discussion}
\label{sec:discussion}

In the UrQMD calculations using the reaction of $^6$Li+$^{12}$C at 2 $A$ GeV, the $\Lambda$-hypernucleus of interest in the target-rapidity region produced by single-charge exchange with $\Lambda K^+$ production is the $\SH{Be}{12}$ hypernucleus converted from the $^{12}$C target nucleus, which is shown at $(A',Z')=(A_{\rm target},Z_{\rm target}-1-1)=(12,6-1-1) = (12,4)$ in panel (c) in Fig. \ref{fig:AZ_all}. The first `$-1$' in the second pair of parentheses corresponds to the charge exchange from a proton to a neutron, while the second instance is caused by $\Lambda K^+$ production from a proton. The $\SH{Be}{12}$ hypernucleus should be produced by peripheral collisions with an outgoing projectile residue with the same mass number as the $^6$Li projectile, but with an increased charge number $Z$ based on the single change from a neutron to a proton, which corresponds to $(A',Z')=(A_{\rm projectile}, Z_{\rm projectile}+1) = (6,3+1) = (6,4)$ in panel (b) in Fig. \ref{fig:AZ_all}. For the double-charge-exchange reaction with $\Lambda K^+$ production, the hypernucleus of interest produced in the target rapidity region is the $\SH{Li}{12}$ hypernucleus, which is shown as $(A',Z')=(A_{\rm target},Z_{\rm target}-2-1)=(12,6-2-1) = (12,3 )$ in panel (c) in Fig.  \ref{fig:AZ_all}, where the associated projectile residue at $(A',Z')=(A_{\rm projectile}, Z_{\rm projectile}+2)=(6,3+2) = (6,5)$ is shown in panel (b) in the same figure. 

\begin{figure*}[htb]
\centering
  \resizebox{120mm}{!}{\includegraphics{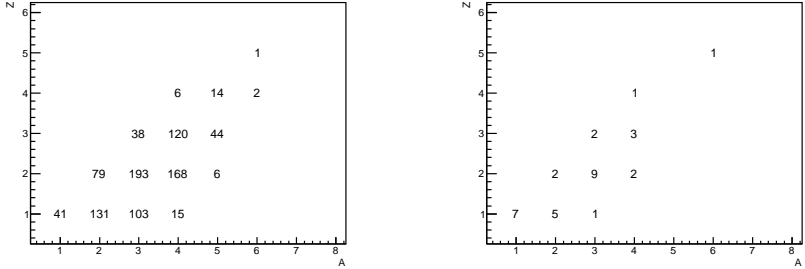}} 
\caption{Correlations between the mass number ($A$) and charge ($Z$) for projectile residues with the production of $\SH{Be}{12}$ (left panel) and $\SH{Li}{12}$ (right panel) in the target-rapidity region. Numbers correspond to counts for each pair of $(A,Z)$. }
\label{fig:pro_res}
\end{figure*}

\begin{table}[htb]
\caption{Summary of the cross section, production, and expected observation of the $^{12}$C($^{6}$Li, $^{6}$Be $K^+$)$\SH{Be}{12}$ and $^{12}$C($^{6}$Li, $^{6}$B $K^+$)$\SH{Li}{12}$ reactions. See the text for details. }
\label{tbl:crosssection}
\begin{center}
\begin{tabular}{ c | c | c | c | c }
\hline \hline
 Reaction& Cross section  & \multicolumn{2}{c|}{Production}   & Observation\\
 & & /day & /week & /week \\
 \hline
 $^{12}$C($^{6}$Li, $^{6}$Be $K^+$)$\SH{Be}{12}$ & $0.05$ nb & $1.7\times 10^2$ & $1.2\times 10^3$ & $2.4\times 10^2$\\
 $^{12}$C($^{6}$Li, $^{6}$B $K^+$)$\SH{Li}{12}$ & $0.03$ nb& $1.0\times 10^2$& $7.0\times 10^2$ & $1.5\times 10^2$\\
 \hline \hline
\end{tabular}
\end{center}
\end{table}

Fig. \ref{fig:pro_res} presents the $A$-$Z$ correlations of projectile residues with the production of $\SH{Be}{12}$ (left panel) and $\SH{Li}{12}$ (right panel) in the target-rapidity region. Both $\NH{Be}{6}$ at $(A,Z)=(6,4 )$ in the left panel and $\NH{B}{6}$ at $(A,Z)=(6,5 )$ in the right panel are produced, but with very small statistics (i.e., two counts for $\NH{Be}{6}$ and one count for $\NH{B}{6}$). Even with these small statistics, one can estimate the production cross section values of the $\SH{Be}{12}$ and $\SH{Li}{12}$ hypernuclei via the single- and double-charge-exchange reactions with $\Lambda K^+$ production, namely $^{12}$C($^{6}$Li, $^{6}$Be $K^+$)$\SH{Be}{12}$ and $^{12}$C($^{6}$Li, $^{6}$B $K^+$)$\SH{Li}{12}$, respectively, by using Eq. (\ref{eqn:eqn1}).  The derived cross section values are $0.05 \pm 0.035$ nb and $0.03 \pm 0.03$ nb, respectively, and the central values are shown in Table \ref{tbl:crosssection}. 
By taking the central values of the derived cross sections with a beam intensity of $10^8$ particles per second and target thickness of 8 g/cm$^2$, which is the same thickness employed in the HypHI Phase 0 experiment \cite{RAPPOLD2013170}, the production rates per day and per week can be calculated. These values are also provided in the table.    
It should be noted that they correspond to the proposed production reactions of ($^{12}$C, $^{12}$N $K^+$) and ($^{9}$Be, $^{9}$C $K^+$), respectively. 

\begin{figure*}[htb]
\centering
  \resizebox{120mm}{!}{\includegraphics{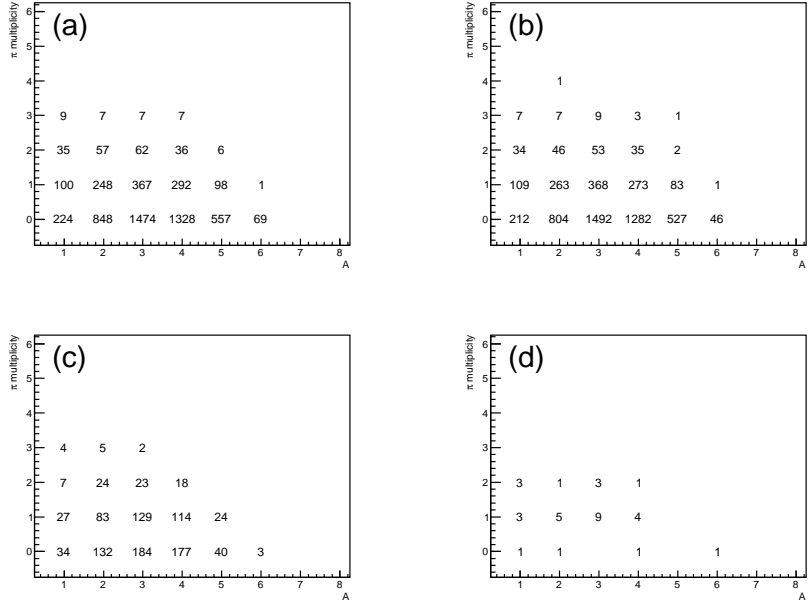}} 
\caption{(a) Multiplicity of $\pi$-mesons as a function of the mass number ($A$) of the projectile residues for the production of $\SH{C}{12}$, (b) $\SH{B}{12}$, (c) $\SH{Be}{12}$, and (d) $\SH{Li}{12}$. The numbers correspond to the counts for each $(A,\pi \: {\rm multiplicity})$ pair.}
\label{fig:mul_pi}
\end{figure*}

Because the hypernuclear production cross section in the proposed novel reaction can be very small compared to those in other reaction channels, the background for observation must be carefully evaluated. In the proposed method, the hypernuclei of interest are observed using the missing mass method by measuring the four-momentum of an incident projectile, $K^+$, and an outgoing projectile residue. If the production of $\pi$-mesons is associated with hypernuclear production events and if the $\pi$-mesons are not measured, then those events are attributed to the background. Therefore, the multiplicity of $\pi$-mesons per event in the proposed reaction is crucial. Because the production of $\SH{Be}{12}$ and $\SH{Li}{12}$ is too small to study the multiplicity of $\pi$-mesons, the $\pi$-meson multiplicity associated with the production of $\SH{C}{12}$ and $\SH{B}{12}$ in the target-rapidity region was first investigated to study the trends of $\pi$-meson production in peripheral collisions. Similar to the proposed production reactions, these hypernuclei are produced by the following peripheral collisions:
\begin{itemize}
\item[(i)] for $\SH{C}{12}$, the no-charge-exchange reaction ($^{6}$Li, $^{6}$Li $K^0$) with 1.2 nb and single-charge-exchange reaction ($^{6}$Li, $^{6}$He $K^+$) with 0.6 nb, 
\item[(ii)] for $\SH{B}{12}$, the no-charge-exchange reaction ($^{6}$Li, $^{6}$Li $K^+$) with 0.2 nb and single-charge-exchange reaction ($^{6}$Li, $^{6}$Be $K^0$) with 1.0 nb. 
\end{itemize}
Panels (a) and (b) in Fig. \ref{fig:mul_pi} present the multiplicity of $\pi$-mesons as a function of the mass number $A$ of the projectile residues for cases (i) and (ii), respectively. One can clearly see that the $\pi$-meson multiplicity is dominated at zero by the associated projectile residues with $A=6$ corresponding to peripheral collisions. For case (i) in panel (a), the ratio of multiplicity = 0 to multiplicity = 1 at A = 6 is 69 / 1. For case (ii) in panel (b), this ratio is 46 / 1. The amount of non-zero multiplicity increases as $A$ decreases because the centrality of the collisions increases. These results indicate that the number of event reconstructions with missing $\pi$-mesons is sufficiently small when we set the mass number of the outgoing projectile to be the same as that of the incident projectile. Additionally, those events do not contribute significantly to the background. Although the statistics are small for the production of $\SH{Be}{12}$ and $\SH{Li}{12}$, the multiplicity distributions of $\pi$-mesons were verified, as shown in panels (c) and (d) in Fig.  \ref{fig:mul_pi}. Similar trends can be observed and only a null multiplicity for $A=6$ can be observed for both cases with small statistics.

\begin{figure*}[htb]
\centering
  \resizebox{120mm}{!}{\includegraphics{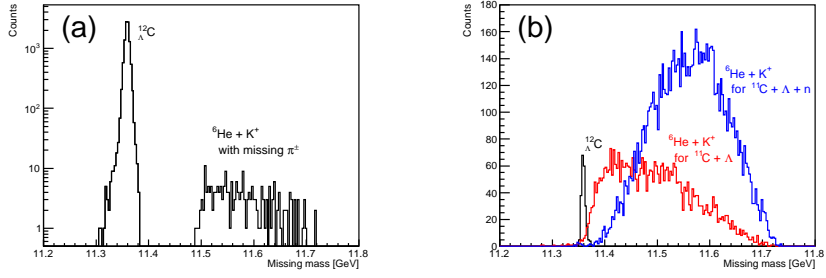}} 
\caption{Left panel (a): missing mass distributions reconstructed by measurements of $^6$He and $K^+$ for the perfect case, $^6$Li + $^{12}$C at 2 $A$ GeV $\rightarrow $ $^6$He + $K^+$ + $\SH{C}{12}$, and an incomplete case with missing $\pi ^+$, $^6$Li + $^{12}$C at 2 $A$ GeV $\rightarrow $ $^6$He + $K^+$ + $\SH{B}{12}$ + $\pi ^+$. Right  panel (b): Missing mass distributions based on observing $^6$He and $K^+$ for the case of $^6$Li + $^{12}$C at 2 $A$ GeV $\rightarrow $ $^6$He + $K^+$ + $^{11}$C + $\Lambda $ (red colour) and $^6$Li + $^{12}$C at 2 $A$ GeV $\rightarrow $ $^6$He + $K^+$ + $^{10}$C + n + $\Lambda $ (blue colour) with the $\SH{C}{12}$ peak. The width of the $\SH{C}{12}$ peak is approximately 4.5 MeV in $\sigma $. }
\label{fig:missingmass}
\end{figure*}

Although the $\pi$-meson multiplicity is dominated at zero in peripheral collisions with the mass number of projectile residues being equal to that of the projectiles, the contributions of the background with events missing the detection of $\pi$-mesons were investigated in preparation for the worst-case scenario. 
In this study, the channel of interest (perfect case) is \\
$^6$Li + $^{12}$C at 2 $A$ GeV $\rightarrow $ $^6$He + $K^+$ + $\SH{C}{12}$ \\
for producing the $\SH{C}{12}$ hypernucleus. 
Phase-space calculations were performed for this channel and kinematic cut conditions were applied to the particles and nuclei produced by this reaction to reproduce the kinematics in our UrQMD calculations. 
The $\SH{C}{12}$ hypernucleus was observed using the missing mass method by measuring $^6$He and $K^+$. We assume that the projectile residue is measured by a high-resolution forward spectrometer such as the Super-FRS facility at FAIR or HFRS facility at HIAF. Therefore, the momentum resolution $\Delta p/p$ for measurements of $^6$He is set to $10^{-4}$. The momentum resolution for $K^+$ was assumed to be $10^{-2}$. A sharp peak on the left side of panel (a) in Fig. \ref{fig:missingmass} corresponds to the missing mass peak of $\SH{C}{12}$ and its width is approximately 4.5 MeV in $\sigma$.  
The corresponding background channel is\\
$^6$Li + $^{12}$C at 2 $A$ GeV $\rightarrow $ $^6$He + $K^+$ + $\SH{B}{12}$ + $\pi ^+$, \\
where similar phase-space calculations were performed. For this channel, the missing mass is calculated by measuring $^6$He and $K^+$ while missing the $\pi ^+$-mesons, which contribute to the background. 
To estimate the amount of background, the ratio of the signal to the background was derived from the results of the UrQMD calculations. This ratio is $46/1$, which comes from the values at $A=6$ in panel (b) in Fig. \ref{fig:mul_pi}. The wide distribution on the right side in panel (a) in Fig. \ref{fig:missingmass} corresponds to the background distribution with missing $\pi ^+$-mesons.  The background distribution is clearly separated from the target missing mass peak.

\begin{figure*}[htb!]
\centering
  \resizebox{120mm}{!}{\includegraphics{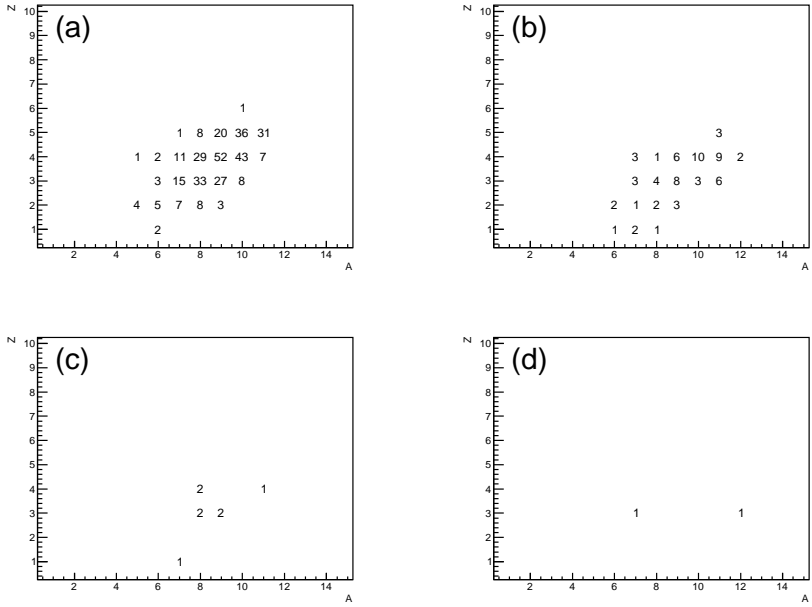}} 
\caption{Correlations between the mass number ($A$) and charge ($Z$) of products in the target-rapidity region produced by the $^{12}$C($^{6}$Li, $^{6}$Be $K^+$) reaction. Panels (a) and (b) present nuclear fragments and hypernuclei, respectively. Panels (c) and (d) present similar distributions for nuclear fragments and hypernuclei, respectively, but for the $^{12}$C($^{6}$Li, $^{6}$B $K^+$) reaction. Numbers correspond to the counts for each pair of $(A,Z)$.}
\label{fig:AZ_6Be6BK}
\end{figure*}

The other major contributions to the background could be products other than the hypernuclei of interest in the target-rapidity region, even with the perfect selection of projectile residues in the reactions of $^{12}$C($^{6}$Li, $^{6}$Be $K^+$) with single-charge exchange and $^{12}$C($^{6}$Li, $^{6}$B $K^+$) with double-charge exchange combined with a coincident measurement of $K^+$. Fig. \ref{fig:AZ_6Be6BK} presents the $A$-$Z$ distributions of products in the target-rapidity region with the production of projectile residues ($^{6}$Be) and $K^+$ for nuclear fragments (panel (a)) and hypernuclei (panel (b)) for the $^{12}$C($^{6}$Li, $^{6}$Be $K^+$) reaction. One can see that there are a variety of nuclear fragments and hypernuclei produced in the target-rapidity region other than the hypernucleus of interest, namely $\SH{Be}{12}$, all of which contribute to the production of background. 
Similar distributions for the $^{12}$C($^{6}$Li, $^{6}$B $K^+$) reactions can be observed in panels (c) and (d). The ratios of the hypernuclei of interest to the total of the other products (nuclear fragments and other hypernuclei) with the perfect detection of $K^+$ and the projectile residues of interest are $2/425\approx 4.7\times 10^{-3}$ and $1/9\approx1.1\times 10^{-1}$ for $^{12}$C($^{6}$Li, $^{6}$Be $K^+$), and $^{12}$C($^{6}$Li, $^{6}$B $K^+$), respectively. It should be noted that the signal-to-background ratio is better for the $^{12}$C($^{6}$Li, $^{6}$B $K^+$) reaction because the products contributing to the background are also hardly produced in the double-charge-exchange reaction. 

Based on the poor signal-to-background ratios of these reactions, the observation of hypernuclei of interest depends on how the background is distributed around the missing mass peaks. Therefore, we consider the following two channels for producing major backgrounds:  \\
$^6$Li + $^{12}$C at 2 $A$ GeV $\rightarrow $ $^6$He + $K^+$ + $^{11}$C + $\Lambda $ \\ 
and \\
$^6$Li + $^{12}$C at 2 $A$ GeV $\rightarrow $ $^6$He + $K^+$ + $^{10}$C + n + $\Lambda $. \\ 
The background attributed to the first reaction with a dissociation of one $\Lambda$-hyperon is the most crucial because the probability of this process is expected to be large and because the background distribution should be closer to the signal peak in the missing mass distributions. The latter reaction is also important, but it is expected that the background distributions will be further away from the signal because there are two dissociated particles ($\Lambda $ and n). 
The missing mass distributions for these two channels when measuring $^6$He and $K^+$ are reconstructed as the background in panel (b) in Fig.  \ref{fig:missingmass} for the case of $^6$He + $K^+$ + $^{11}$C + $\Lambda $ with red colour and for the case of $^6$He + $K^+$ + $^{10}$C + n + $\Lambda $ with blue colour. We also present the sharp distribution of the reconstructed $\SH{C}{12}$. Signal-to-noise ratios were estimated from Fig. \ref{fig:AZ_6Be6BK} and we overestimated the background to consider the worst-case scenario by summing the number of products over the mass number $A$ of the produced nuclei in both channels. The estimated signal-to-background ratios are $2/(37+1)=2/38\approx  5.3\times 10^{-2}$ for the case of $^6$He + $K^+$ + $^{11}$C + $\Lambda $ and $2/(1+36+43+8)=2/88\approx 2.3\times 10^{-2}$ for the case of $^6$He + $K^+$ + $^{10}$C + n + $\Lambda $. These ratios are considered for the distributions in panel (b) in Fig. \ref{fig:missingmass}. The background distribution of the first channel, namely $^6$He + $K^+$ + $^{11}$C + $\Lambda $, is closer to the signal. However, a peak can be clearly observed. The background distribution of the latter channel, namely $^6$He + $K^+$ + $^{10}$C + n + $\Lambda $, is larger. However, it is also further away from the signal peak. These background distributions clearly exhibit the expected trend of moving further away from the missing mass peak as more particles are dissociated. 
Therefore, the background distributions are attributed to the other nuclear and hypernuclear products with more dissociated particles, as shown in Fig. \ref{fig:AZ_6Be6BK}, which are even further away from the missing mass peak and do not disturb the observation of the peak.  

In our experiments, projectile residues were measured by a high-momentum-resolution forward spectrometer, such as the Super-FRS facility \cite{Super-FRS} at FAIR \cite{FAIR} or HFRS facility at HIAF \cite{HIAF}, and the typical acceptance for projectile residues in the conseidered case is approximately 40\%. We are currently designing a detector system to measure charged $K^+$ and $\pi$-mesons. The estimated efficiency for $K^+$ produced in the reactions of interest in this work is approximately 50\%. These values combined with the estimated production rates in Table \ref{tbl:crosssection} indicate that the expected numbers of observations for the hypernuclei of interest are $2.4\times 10^2$ per week for $^{12}$C($^{6}$Li, $^{6}$Be $K^+$)$\SH{Be}{12}$ and $1.5\times 10^2$ per week for  $^{12}$C($^{6}$Li, $^{6}$B $K^+$)$\SH{Li}{12}$. These results are summarised in Table \ref{tbl:crosssection}. Similar observed counts are also expected for the reactions proposed in this work, namely ($^{12}$C, $^{12}$N $K^+$), and ($^{9}$Be, $^{9}$C $K^+$). 
One can see reasonable separations between the signal peaks and broad background distributions in Fig. \ref{fig:missingmass}. The missing mass peaks of the hypernuclei of interest can be observed through measurements over one week, although the signal-to-background ratios are small. These observed counts could be sufficient for studying extremely neutron-rich hypernuclei, as shown in Table \ref{tbl:list}. It should be noted that the expected observed counts for hyper-resonance states may be inflated because the cross sections producing those resonance states are expected to be larger than those for the associated hypernuclei.

%%%%%%%%%%%%%%%%%%%%%%%%%%%%%%%%%%%%%%%%%%%%%%%%%%%%%%%%%%%%%%
\section{Summary}

We proposed a novel method for producing very-neutron-rich hypernuclei, as well as hyper-resonance states, in the target-rapidity region by employing single- and double-charge-exchange reactions of heavy ion beams combined with $\Lambda K^+$ production. The hypernuclei of interest were observed using the missing mass method by measuring an outgoing projectile residue and $K^+$ simultaneously. Outgoing projectile residues should be bound only by their ground states. Therefore, we selected two reaction channels of pp($^{12}$C, $^{12}$N $K^+$)n$\Lambda$ with single-charge exchange and ppp($^{9}$Be, $^{9}$C $K^+$)nn$\Lambda$ with double-charge exchange. The corresponding conversion processes take place inside the target nucleus.  Based on these reactions, as shown in Table \ref{tbl:list}, a variety of very-neutron-rich $\Lambda$-hypernuclei can be produced, including $Z=0$ bound states and resonances. To produce hypernuclei using these reactions, the energy should be greater than or equal to 2 $A$ GeV to produce a sufficient amount of hyperons. However, data on charge-exchange reactions at such high energies are scarce. 

To study the feasibility of the proposed method, we employed a microscopic transport model called UrQMD. Nuclear fragments and hypernuclei were produced by applying kinematic cut conditions to the events produced by UrQMD. We chose the reaction of $^6$Li+$^{12}$C at 2 $A$ GeV for our studies using UrQMD because the calculations with this reaction can reproduce the hypernuclear and $\Lambda$ production cross sections observed in the HypHI Phase 0 experiment and because we already have approximately 15 billion events for this reaction. Among these events, single-charge- and double-charge-exchange reactions with $\Lambda K^+$ production were investigated via the reaction channels of ($^{6}$Li, $^{6}$Be $K^+$) and ($^{6}$Li, $^{6}$B $K^+$). Production cross sections, estimated yields, and the expected amount of reconstruction using the missing mass method were evaluated for very-neutron-rich hypernuclei produced by the $^{12}$C($^{6}$Li, $^{6}$Be $K^+$)$\SH{Be}{12}$ and  $^{12}$C($^{6}$Li, $^{6}$B $K^+$)$\SH{Li}{12}$ reactions. Additionally, the backgrounds associated with these reactions were investigated. It was shown that very-neutron-rich hypernuclei can be produced by single- and double-charge-exchange reactions with $\Lambda K^+$ production and that they can be observed using the missing mass method because the missing mass signals are well separated from the large background distributions, although the estimated statistics are not large. 

This work demonstrated that the proposed method can open new avenues for studying very-neutron-rich hypernuclei and hyper-resonance states in the Super-FRS facility at FAIR and HFRS facility at HIAF. 

%%%%%%%%%%%%%%%%%%%%%%%%%%%%%%%%%%%%%%%%%%%%%%%%%%%%%%%%%%%%%%
\section*{Acknowledgement}
The authors thank H.J. Ong of the Institute of Modern Physics and S. Terashima of Beihang University for their fruitful discussions of charge-exchange reactions. The authors also thank J. Yoshida of RIKEN and Tohoku University for commenting on the manuscript.

%\begin{acknowledgements}
%\end{acknowledgements}



\end{document}